\documentclass[onecolumn,amsmath,amssymb,floatfix,nofootinbib]{revtex4-2}

\usepackage[utf8]{inputenc}
\usepackage{amsfonts,amsmath,amssymb}
\usepackage{hyperref}
\usepackage{graphicx}
\usepackage{multirow}
\usepackage{verbatim}
\usepackage{bm}
\usepackage[textsize=scriptsize,backgroundcolor=red!70,linecolor=red]{todonotes}
\usepackage{paralist}
\usepackage{dcolumn}
\usepackage{latexsym}

\usepackage{natbib}

\newcommand{\ii}{\mathrm{i}}
\newcommand{\ee}{\mathrm{e}}

\newcommand{\abs}[1]{\big\lvert #1 \big\rvert}

\newcommand{\ie}{i.e.}

\newcommand{\ud}{\ensuremath{\mathrm{d}}}

\newcommand{\chisq}{\ensuremath{\chi^{2}}}

\newcommand{\porpb}{\ensuremath{\overset{\scriptscriptstyle \left(-\right)}{p}}}
\newcommand{\fform}{F}
\newcommand{\bvec}[1]{\ensuremath{\bm #1}}
\newcommand{\dof}{\ensuremath{\text{d.o.f}}}
\newcommand{\apr}{\ensuremath{\alpha^\prime}}

\begin{document}
\title{The proton inelastic cross section at ultrahigh energies}
 
\author{Atri Bhattacharya}
\email{a.bhattacharya@uliege.be}
\affiliation{Space sciences, Technologies and Astrophysics Research (STAR) Institute,
             Université de Liège, Bât.~B5a, 4000 Liège,
             Belgium}

\author{Jean-René Cudell}
\email{jr.cudell@uliege.be}
\affiliation{Space sciences, Technologies and Astrophysics Research (STAR) Institute,
             Université de Liège, Bât.~B5a, 4000 Liège,
             Belgium}

\author{Rami Oueslati} 
\email{rami.oueslati@uliege.be}
\affiliation{Space sciences, Technologies and Astrophysics Research (STAR) Institute,
             Université de Liège, Bât.~B5a, 4000 Liège,
             Belgium}

\author{Arno Vanthieghem}
\email{vanthieg@slac.stanford.edu}
\affiliation{High Energy Density Science Division (HEDS), 
SLAC National Accelerator Laboratory, Menlo Park, California 94025, USA}

\date{\today}

\begin{abstract}
We study the consequences of high-energy collider data on the best fits to
total, elastic, and inelastic cross sections for $pp$ and $p\bar{p}$ scattering
using two very distinct unitarisation schemes: the eikonal and the $U$-matrix.
Despite their analytic differences, we find that the two schemes lead to almost identical 
predictions up to EeV energies, with differences only becoming significant at
GUT-scale and higher energies.
\end{abstract}

\maketitle

Man-made accelerators and indirect detection methods of high-energy cosmic rays such as extensive air showers, at the core of high-energy and multi-messenger astrophysics, have drawn a particular attention to the modeling of the high-energy hadronic interactions. A comprehensive treatment of the $pp$ and $p\bar{p}$ cross sections with quantum chromodynamics being elusive for the moment, one has to rely on some generic arguments about unitarity and analyticity of the scattering matrix to derive phenomenological estimates of the high-energy total, elastic and inelastic cross sections. In that regard, experimental studies, most notably those related to cosmic-ray showers, often use the 2002 fits to the total cross section that successfully predicted the LHC $pp$ total cross section \citep{Cudell:2002xe}.
Besides the fact that there are a lot of relevant data that have since appeared \citep{Antchev:2011vs,Antchev:2013gaa,Antchev:2013iaa,Antchev:2013paa,Antchev:2017dia,Aad:2014dca,Aaboud:2016ijx,Aad:2011eu,Myska:2017iqc,Aaij:2018okq, Abelev:2012sea,Antchev:2013haa}, these fits have the drawback that they cannot self-consistently relate the total cross section to the elastic and inelastic ones.
Since the inelastic cross section is key to computing multiple minijet production from cosmic-ray interactions with the atmosphere at ultra-high energies, the relation between the total and inelastic cross sections is therefore essential to the description of extensive air showers.
It is at the core of hadronic interaction models adopted in Monte Carlo event generators such as \texttt{SIBYLL} \cite{Engel:2019dsg} and \texttt{QGSJET} \cite{Ostapchenko:2019few}.

In this letter, we want to address this
problem\footnote{The question of  the very forward component of the showers, which is linked to the diffractive cross section, will be considered in a separate paper.}. In order to relate elastic, inelastic, and total cross sections, one needs a physics model of the elastic amplitude. This is typically made of two ingredients: an  elastic amplitude at the Born level, which encapsulates the  elementary exchange (and can be extracted from low-energy data),
and a scheme that takes into account multiple exchanges, which become increasingly important at higher energies and without which the elastic amplitude would exceed the unitarity limit.

The Born term of interest corresponds to pomeron exchange, and is reasonably constrained. We normalize the elastic amplitude $a(s,t)$ so that the differential cross section for elastic scattering  is written as
\newcommand{\bq}{\bvec{q}}
\newcommand{\bb}{\bvec{b}}
\begin{equation}
	\frac{\ud\sigma_{el}}{\ud t} = \frac{\abs{a(s,t)}^2}{16\pi s^2}\,,
	\label{eq:dsigmadt}
\end{equation}
where $t = -\bq^2$ is the square of the momentum transfer.
The Born term can then be written 
using the pomeron trajectory $\alpha(t)$, 
the proton elastic form factor $F_1(t)$ and the coupling pomeron-proton $g_p$, as
\begin{equation}
	a(s,t) =g_p^2\, \fform_1(t)^2 \left( \frac{s}{s_0} \right)^{\alpha(t)}\, \xi(t)\,,
	\label{eq:amp_t}
\end{equation}
with $\xi(t)$ the signature factor
\begin{equation}
	\xi(t) =-e^{-i\pi\alpha(t)\over 2}\,.
	\label{eq:amp_t2}
\end{equation}
The pomeron trajectory is close to a straight line \cite{Cudell:2005sg}, and we take it to be $\alpha(t)=1+\epsilon+\alpha' t$.

At high energy, the growth of this pomeron amplitude and eventual violation of unitarity is most
clearly seen in the impact-parameter representation, where 
the Fourier transform of the amplitude $a\left( s, t \right)$ rescaled by $2s$ is equivalent to a partial wave
\begin{equation}
	\chi(s, \bvec{b}) = \int \frac{\ud^2\bvec{q}}{\left( 2\pi \right)^2}
	\frac{a(s,t)}{2s} \ee^{\ii \bvec{q}\cdot \bvec{b}}.
	\label{eq:Gsb}
\end{equation}
 The norm of the partial wave signals two
important regimes. When it reaches unity, around $\sqrt{s}=2$ TeV \cite{Cudell:2003dz}, the model enters the black-disk regime -- \ie\ maximum inelasticity. When it reaches two, the model begins to violate unitarity. Both regimes start at small impact parameter and spread to higher values of $b$, and signal that  multiple exchanges have to be taken into account.

It is thus
necessary to introduce unitarization schemes that take into account multiple scatterings by mapping
the amplitude $\chi(s,\bb)$  to the physical amplitude $X(s,\bb)$. The latter reduces to $\chi(s,\bb)$ for small $s$, is confined to the unitarity circle $\abs{X(s, \bvec{b})-i} \leqslant 1 $, and bears the same relation as Eq.(\ref{eq:Gsb}), but this time to the unitarized amplitude $A(s,t)$:
 \begin{equation}
	X(s, \bvec{b}) = \int \frac{\ud^2\bvec{q}}{\left( 2\pi \right)^2}
	\frac{A(s,t)}{2s} \ee^{\ii \bvec{q}\cdot \bvec{b}}.
	\label{eq:Xsb}
\end{equation}

The eikonal scheme, --- derived for
structureless bodies in optics, potential scattering, and QED --- is
commonly used in the literature.
Another proposed scheme is the $U$-matrix scheme, which can be motivated by a form of
Bethe-Salpeter  equation \cite{Logunov:1971jy}.
Neither of these may be entirely correct in the context of QCD, but going from
one to the other permits an evaluation of the systematics linked to
multiple exchanges.

The eikonal scheme assumes \cite{Cudell:2008yb}:
\begin{equation}
	X_E(s, \bvec b) = \ii \left[ 1 - \ee^{\ii \chi(s, \bvec b)} \right]\,,
	\label{eq:eikx}
\end{equation}
while the $U$-matrix scheme posits:
\begin{equation}
	X_U(s,\bvec b) = \frac{\chi(s, \bvec b)}{1 - \ii \chi(s, \bvec b)/2}.
	\label{eq:umatx}
\end{equation}
In terms of partial waves, the maximum inelasticity is reached in either case for
$ X(s,\bvec{b}) = \ii $, which is also the asymptotic limit of the eikonal scheme at high $s$.

The total and elastic scattering cross sections may be readily expressed in these
representations as 
\begin{align}
	\sigma_{\text{tot}} &= 2 \int\! \ud^{2} \bvec{b}\ \mathfrak{Im} \left( X(s,\bvec{b}) \right),
  &
	\sigma_{\text{el}}  &= \int\! \ud^{2} \bvec{b}\ \abs{X(s, \bvec b)}^{2}.
	\label{eq:sigma}
\end{align}

We shall now use them to fit all the data in $ p\porpb $ scattering above 500 GeV, for which lower trajectories have a 
negligible effect.
\begin{table}[t]
  \begin{tabular}{c@{\hspace{1.5em}}c@{\hspace{1.5em}}r}
    \hline
    \hline\\[-0.8em]
    dataset &number of points &$\chi^2$\\[0.2em]
    \hline\\[-0.7em]
    $\sigma_{tot}$&18  & 21.7\\[0.3em]
    $\sigma_{el}$&11  & 21.3\\[0.3em]
    $\sigma_{in}$&8  & 4.1 \\[0.3em]
    \hline
    \hline
  \end{tabular}
  \caption{\label{tab:data1} The values of $\chi^2$ resulting from independent fits to quadratic polynomials
  in $\log(s)$, illustrating the tensions in some parts of the dataset.}
\end{table}
We obtain 3 distinct datasets (for total, elastic and inelastic cross sections) from the following sources, for a total of  37 data points:
\begin{itemize}
  \item $pp$ total and elastic cross sections from
        TOTEM \cite{Antchev:2011vs,Antchev:2013gaa,Antchev:2013iaa,
        Antchev:2013paa,Antchev:2017dia}, and ATLAS \cite{Aad:2014dca,Aaboud:2016ijx};
  \item $p\bar{p}$ total and elastic cross sections from CDF \cite{Abe:1993xx},
        E710 \cite{Amos:1990jh,Amos:1992zn}, and E811
        \cite{Avila:1998ej,Avila:2002bp} experiments at TeVatron; and UA4 at
         S$\rm p\bar p$S \cite{Bozzo:1984rk};
  \item Direct measurements of inelastic cross sections, \ie\ not derived from
	      total and elastic measurements, from UA5 at  S$\rm p\bar p$S
	      \cite{Alner:1986iy}, ATLAS \cite{Aad:2011eu,Myska:2017iqc}, LHCb
	      \cite{Aaij:2018okq}, ALICE \cite{Abelev:2012sea}, and TOTEM
        \cite{Antchev:2013haa}.
\end{itemize}

It should be noted that both the total and elastic cross section datasets
include discordant data from different experiments.
This is quantified by a simple consistency check that fits generic
quadratic polynomials in $\log s$ to each dataset and computes the resulting
\chisq.
Table \ref{tab:data1} shows the results with both the elastic and total
cross sections running up $\chisq/\text{d.o.f}$ noticeably greater than 1.
Thus, one obtains a minimum combined $\chi^2$ of 47.1.
This is a well-known problem with these data, first addressed in
\cite{Cudell:1996sh} and later in \cite{Block:2005qm}.
At present, however, the number of data points is simply too small to identify
individual outliers, and hence there is little one can do for lack of better
experimental results.
We shall thus neither filter nor sieve the data, but remember that the best
possible $\chi^2$ is rather high.

\begin{table*}
  \begin{tabular}{l@{\hspace{1.5em}}c@{\hspace{1.5em}}c@{\hspace{1.5em}}c@{\hspace{1.5em}}c@{\hspace{1.5em}}r}
   \hline
   \hline\\[-1em]
   Scheme   & $\epsilon $
            & $\apr $
            & $ g_p $
            & $ t_{0} $
            & $ \frac{\chisq}{\dof} $\\[0.3em]
            \hline\\[-0.6em]
   Eikonal
            & $ 0.11 \pm 0.01 $
            & $ 0.31 \pm 0.19 $
            & $ 7.3  \pm 0.9  $
            & $ 1.9  \pm 0.4  $
            & $ 1.442 $ \\[0.5em]
   U-matrix
            & $ 0.10 \pm 0.01 $
            & $ 0.37 \pm 0.28 $
            & $ 7.5  \pm 0.8 $
            & $ 2.5  \pm 0.6  $
            & $ 1.436 $ \\[0.5em]
   
   \hline
   \hline
\end{tabular}

  \caption{\label{tab:fits1}\chisq/\dof\ and best-fit parameters obtained using the
           eikonal and $U$-matrix unitarisation schemes.
          }
\end{table*}

\begin{figure}[h]
\includegraphics[width=0.65\textwidth]{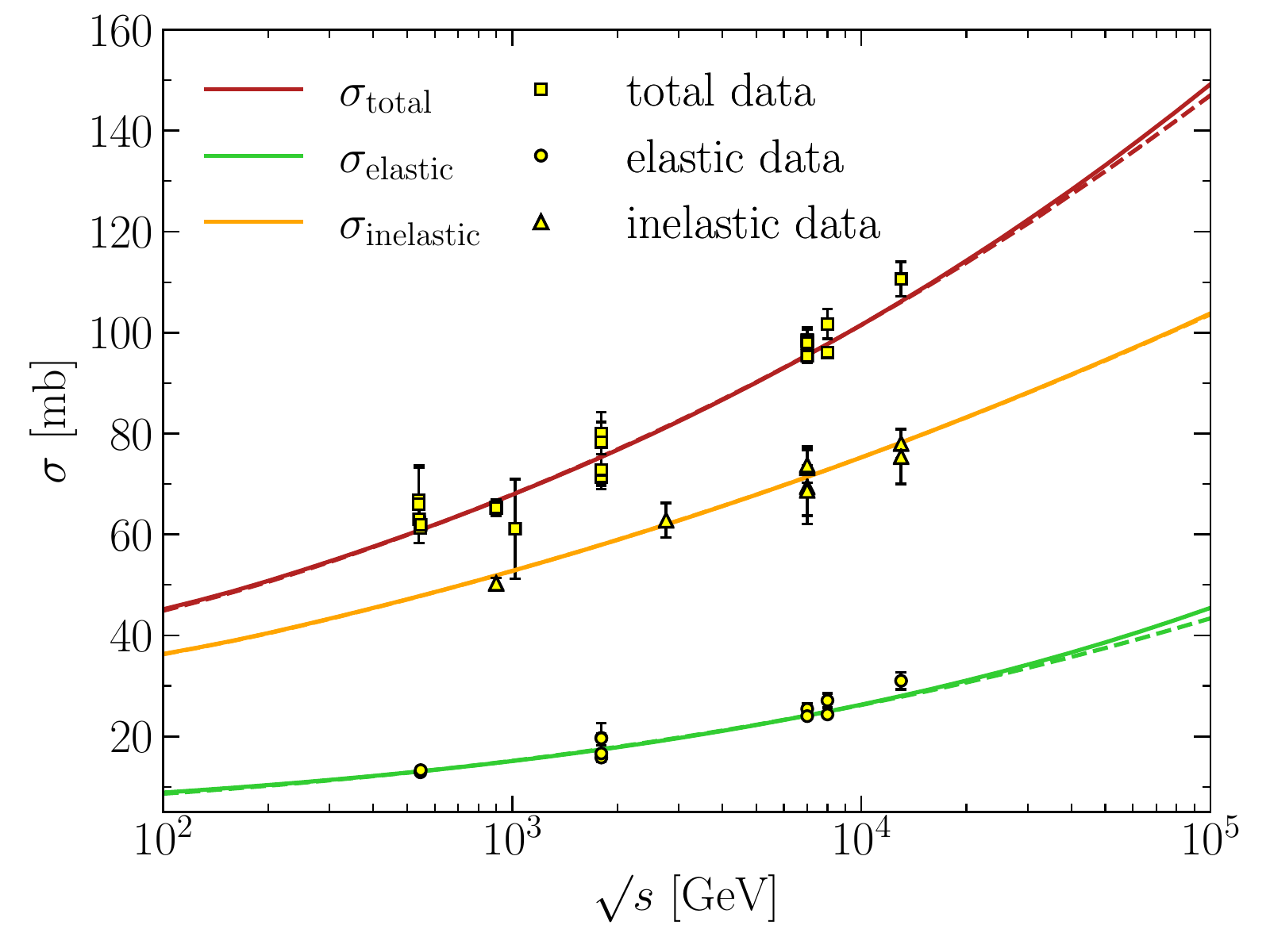}
\caption{\label{fig:ndsec}Total, elastic and inelastic cross sections obtained with best-fit
         parameters for the $U$-matrix scheme (solid curves) and the eikonal scheme
         (dashed curves).}
\end{figure}

We use a dipole-like form factor for the proton $F_1 = 1/(1-t/t_0)^2$.
The parameters in our fit thus include $\epsilon$ and $\alpha^\prime$ describing
the pomeron trajectory, the coupling constant $g_p$, and finally the form-factor
scale $t_0$.

The results of our fits using either unitarisation scheme are shown
in Table \ref{tab:fits1} and in Fig.~\ref{fig:ndsec}.
We obtain $ \chisq / \dof = 1.436\ (1.442) $ when using the $U$-matrix (eikonal)
scheme.
Note that, although at face value the fit obtained using either
scheme only has a seemingly poor $ \chisq / \dof $, the value of the total
$\chisq$ --- $47.39\ (47.59)$ for the $U$-matrix (eikonal) scheme --- is very
close to the minimum value --- $47.1$ --- obtained earlier.

These values of the parameters are however quite striking.
Ref. \cite{Cudell:2005sg} managed to disentangle the pomeron contribution at low energy from that of lower-$t$
trajectories, and provided estimates of its coupling, intercept and slope. These values are within $1 \sigma$
of those obtained here for the $U$ matrix, but the eikonal differs significantly from the low-energy results. Hence it seems that for an eikonal scheme, one never recovers the observed one-pomeron simple pole.

Using an exponential form factor \{$F_1 = \exp \left( R_0 t \right)$\},
instead of the dipole form, leads to slightly poorer fits
\{$\chisq/\dof = 1.440\ (1.445) $\}; however, the qualitative picture remains unaltered.
We have also analysed how the fits improve if one uses the generalised eikonal
and $U$-matrix schemes and we find that these generalisations --- at the cost of
an additional free parameter ($ \omega \text{ or } \omega^\prime $) --- do not
improve the fits significantly.

One particular consequence of the relative independence of the elastic cross
section to the choice of the unitarisation scheme is that values of the $ \rho $
parameter remain largely unaffected by the choice of the scheme as well.
We use our best-fit parameters to compute this parameter across different
energies, and find that the corresponding values agree with existing data, except
for the latest TOTEM measurement.
We indeed obtain $ \rho = 0.131 $ at $ \sqrt{s} = 13 $ TeV.  Whether this discrepancy is
due to the fact that we neglect an odderon contribution, or it comes from a problem in the
extraction of $\rho$ from the data \cite{Ezhela:2020hws} is still unclear. As the purpose of this letter
is the evaluation of the inelastic cross section, the exact value of $\rho$ is of little importance
given than it contributes about 1\% to the processes considered here.

\begin{figure}[htb]
    \includegraphics[width=0.65\textwidth]{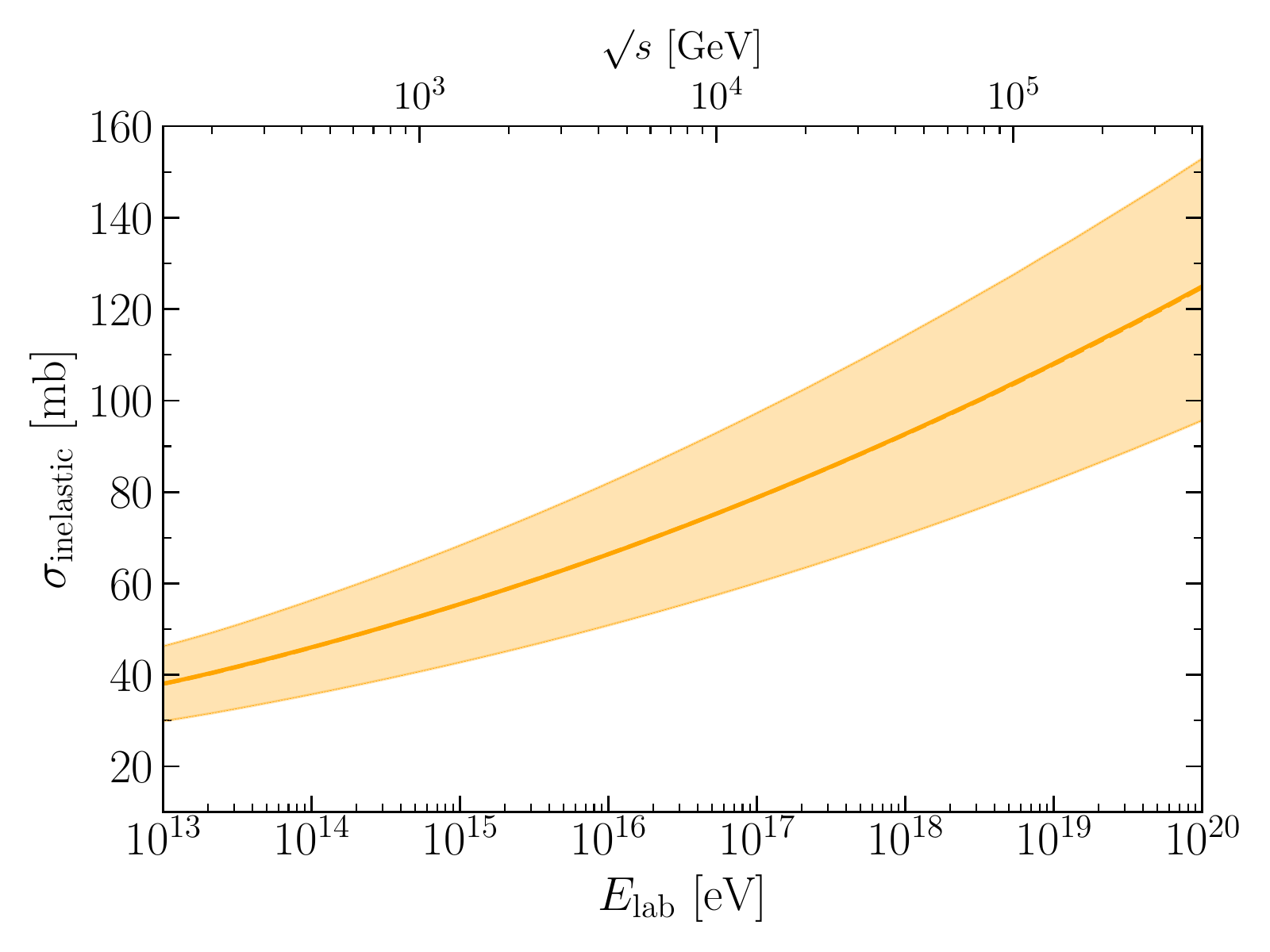}
    \caption{\label{fig:ultrahigh}The 1$\sigma$ band for the inelastic cross section at ultrahigh energies.
    Note that both schemes give almost identical results.}
\end{figure}

We are now in a position to present our results on the inelastic cross section at ultra-high energies.
We obtain them by varying all the parameters of Table~\ref{tab:fits1} in a 1$\sigma$ hyperellipsoid and use
the corresponding curves to evaluate the errors at ultrahigh energies. We show the results in Fig.~\ref{fig:ultrahigh}.
The entwinement of the inelastic cross section with the elastic and total cross sections, which
are much better known, leads to smaller errors than in the case of a fit to inelastic data alone.
Furthermore, despite their very different analytic properties, the two schemes lead to almost
identical predictions.
This gives us confidence that the extrapolation to ultra-high energies is well founded.

\begin{figure}[htb]
    \includegraphics[width=0.65\textwidth]{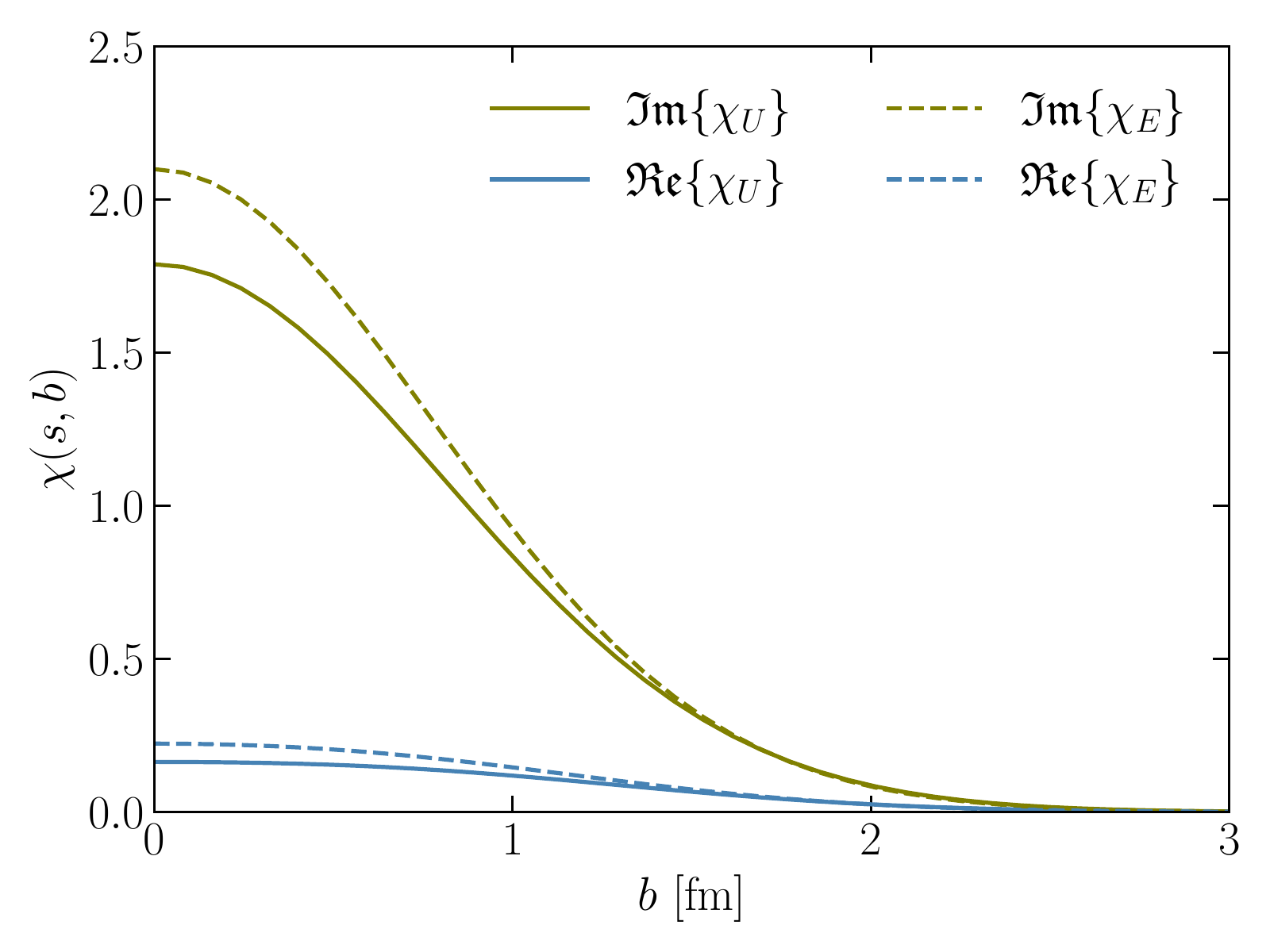}
    \caption{\label{fig:chi}Real and imaginary parts of
             the Born terms $\chi(s,b)$ at $\sqrt{s}=13$ TeV for the $U$-matrix (solid curves) and
             eikonal (dashed curves) schemes.}
\end{figure}

While the inelastic cross sections using either of the two schemes are almost identical,
this alignment happens despite significant differences in the individual order-by-order amplitudes in
the expansion.
We show this for the specific case of the Born term in Fig.~\ref{fig:chi}.
Specifically as it pertains to the inelastic cross section, this order-by-order difference can have major consequences, for example, in Monte Carlo showering codes that depend on the $n$-th term in the Taylor expansion inn $\chi$ to weigh the probability of the $n$ minijets.
In these codes, switching the traditionally used eikonal scheme to
the U-matrix scheme will have an impact on the results, although a full analysis of this impact is beyond the scope of this work.

\begin{figure}[htb]
\centering
\includegraphics[width=0.65\textwidth]{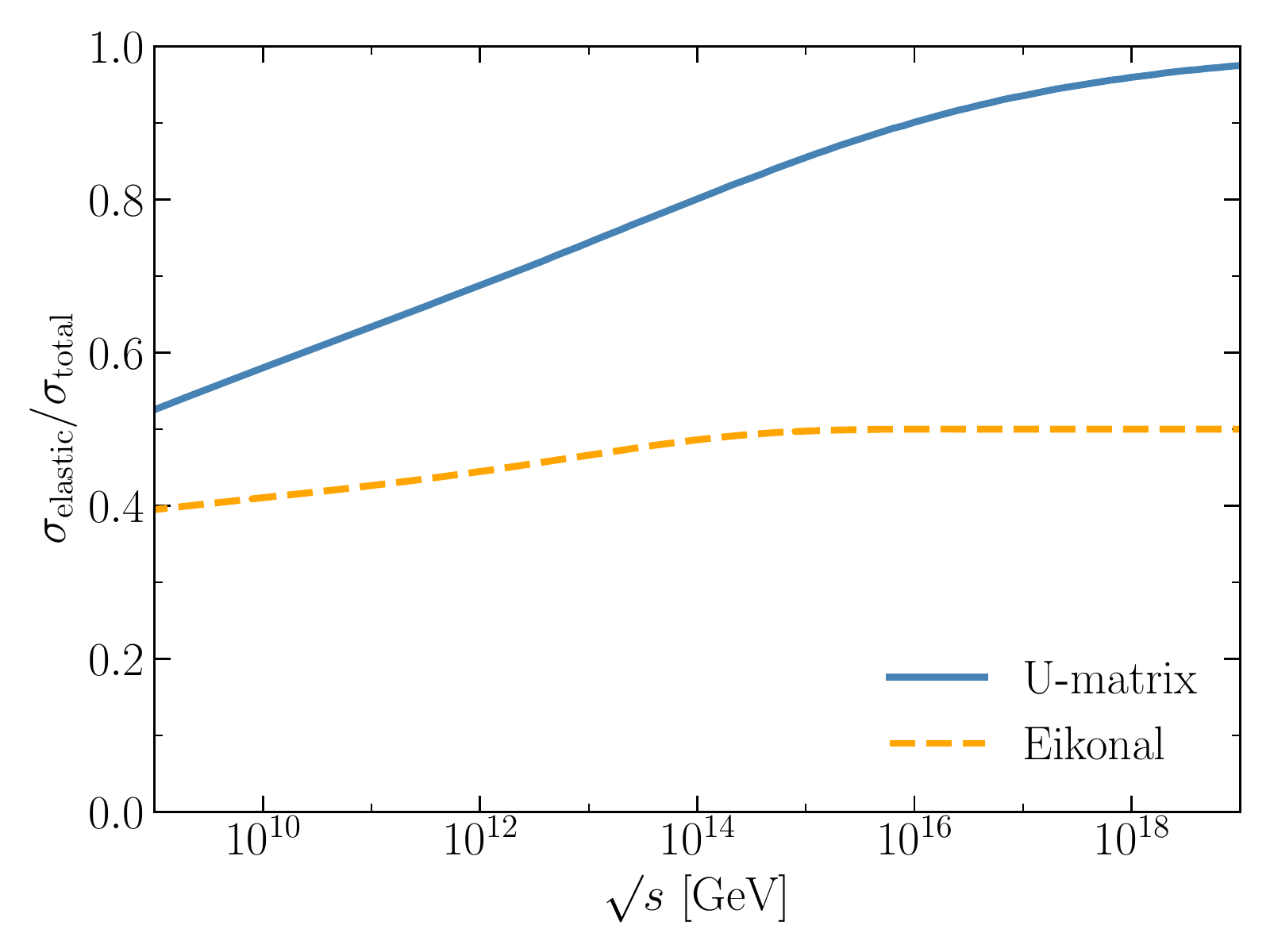}
\caption{\label{fig:ratio-at-gut} The evolution of the ratio of the elastic
         cross section to the total cross section at the GUT-scale and higher energies
         based on the unitarisation scheme chosen.}
\end{figure}

It is important to note that, although we have shown that the total, elastic, and inelastic cross sections obtained using the two schemes remain nearly identical for $\sqrt{s}$ up to tens of TeV, at extremely high $\sqrt{s}$ approaching the grand unification scale
the elastic and inelastic cross sections start differing significantly.
Whereas with the eikonal scheme the elastic cross section reaches parity with
the inelastic cross section at around $ \sqrt{s} = 10^{15} $ GeV and remains so
at higher energies, the $U$-matrix scheme instead predicts continuing growth for the elastic
cross section --- at the cost of the inelastic cross section ---
until it gradually approaches saturation with respect to the total
cross section at some $ \sqrt{s} \gtrsim 10^{19} $ GeV.
This is illustrated in terms of the ratio of the elastic to total cross sections
in Fig.~\ref{fig:ratio-at-gut}.
These extremely high energies are of course beyond the reach of experiments;
such differences are therefore of limited practical relevance.

To summarise, we have used non-diffractive experimental data from
colliders up to $\sqrt{s} = 13$ TeV  to determine the most up-to-date fits
to the total, elastic, and inelastic $p\porpb$ cross sections in the
literature, both for the eikonal and U-matrix unitarisation schemes.
The upshot of our analysis is that the $U$-matrix scheme leads to cross sections that fit the data as well as the eikonal scheme,
which is more conventionally used in most current cosmic-ray Monte Carlo codes.
The corresponding total, elastic, and inelastic cross sections from both schemes are nearly indistinguishable at energies relevant to current and near-future colliders;
they only start showing differences at energies approaching the grand unification scale.
In particular, this allows us to extrapolate the inelastic cross section
up to GZK cut-off energies ($\sim 10^{20}$ eV) uniquely, irrespective
of the unitarisation scheme chosen.
This alignment between the overall inelastic cross sections
notwithstanding, the amplitudes at each order in the series expansions
differ significantly, with potential consequences for Monte Carlo
showering codes.

\acknowledgments{
AB is supported by the  Fonds  de  la  Recherche  Scientifique-FNRS,  Belgium,
under grant No.~4.4503.19.
AB is thankful to the computational resource provided by Consortium des
Équipements de Calcul Intensif (CÉCI), funded by the Fonds de la Recherche
Scientifique de Belgique (F.R.S.-FNRS) under Grant No.~2.5020.11 where a part of
the computation was carried out. This work was also supported by the Fonds de la Recherche Scientifique-FNRS, Belgium, under grant No. 4.4503.19. AV is supported by U.S. DOE Early Career Research Program under FWP100331.
}

\bibliographystyle{unsrtnat}
\bibliography{refs}

\end{document}